MgO barrier-perpendicular magnetic tunnel junctions with CoFe/Pd multilayers and ferromagnetic insertion layers

K. Mizunuma[1, a)], S. Ikeda[1, a)], J. H. Park[1], H. Yamamoto[2,1], H. Gan[1], K. Miura[2,1], H. Hasegawa[1], J. Hayakawa[2], F. Matsukura[1], and H. Ohno[1, b)]

1 *Laboratory for Nanoelectronics and Spintronics, Research Institute of Electrical Communication, Tohoku University, 2-1-1 Katahira, Aoba-ku, Sendai 980-8577, Japan*

2 *Advanced Research Laboratory, Hitachi, Ltd., 1-280 Higashi-koigakubo, Kokubunji-shi, Tokyo 185-8601, Japan*

The authors studied an effect of ferromagnetic ($Co_{20}Fe_{60}B_{20}$ or Fe) layer insertion on tunnel magnetoresistance (TMR) properties of MgO-barrier magnetic tunnel junctions (MTJs) with CoFe/Pd multilayer electrodes. TMR ratio in MTJs with CoFeB/MgO/Fe stack reached 67% at annealing temperature ($T_a$) of 200°C and then decreased rapidly at $T_a$ over 250°C. The degradation of the TMR ratio may be related to crystallization of CoFe(B) into fcc(111) or bcc(011) texture resulting from diffusion of B into Pd layers. MTJs which were *in-situ* annealed at 350°C just after depositing bottom CoFe/Pd multilayer showed TMR ratio of 78% by post annealing at $T_a$ =200°C.

a) Electronic mail (corresponding author): sikeda@riec.tohoku.ac.jp
b) Electronic mail: ohno@riec.tohoku.ac.jp

Spin transfer torque magnetic tunnel junctions with perpendicular magnetic anisotropy electrodes (perpendicular MTJs) attract much interest from the possibility of nonvolatile spin devices compatible with the latest technology node (< 45 nm) in DRAMs having low critical switching current ($I_{c0}$) as well as high thermal stability ($E/k_BT$).[1-4] Currently, perpendicular MTJs prepared using conventional sputtering are actively being studied and have reached the TMR ratios at room temperature (RT) of up to 64% with rare-earth transition metal alloys,[5-7] up to 120% with L1$_0$-ordered (Co, Fe)-Pt,[4,8-10] and up to 15% with Co/(Pd, Pt) multilayers.[11-13] For perpendicular MTJs, $I_{c0}$ and $E/k_BT$ are in a trade-off relation.[1] In order to reduce $I_{c0}$, it is necessary to moderately reduce $E/k_BT$. From this view point, multilayer electrodes have advantages; multilayer is relatively easy to control $M_s$ and $H_k$ by changing the number of the layer stack and the thicknesses. In addition, multilayer films are comparatively easy to realize perpendicular magnetic anisotropy, yet show high magnetic thermal stability. However, it is not clear how one can achieve high tunnel magnetoresistance (TMR) in MTJs based on multilayers. In our previous study,[14] MgO barrier MTJs with $Co_{90}Fe_{10}$/Pd multilayer electrodes showed TMR ratio of a few %. In order to establish the technology for high TMR ratio, we investigated an effect of ferromagnetic ($Co_{20}Fe_{60}B_{20}$ or Fe) layer insertion on TMR properties of MgO-barrier MTJs with $Co_{90}Fe_{10}$/Pd multilayer electrodes.

A set of rf-sputtered MTJ films with a pseudo-spin-valve structure were first studied for electrical measurements. They consist of from the substrate side,

buffer-layer/Pd(1.2)/[Co$_{90}$Fe$_{10}$(0.2)/Pd(1.2)]$_3$/FM(1.8)/MgO(2)/FM(1.8)/[Pd(1.2)/Co$_{90}$Fe$_{10}$(0.2)]$_{10}$/Pd(1.2)/cap-layer (in nm) where FM is Co$_{20}$Fe$_{60}$B$_{20}$ or Fe. For comparison, we fabricated MTJs consisting of buffer-layer/[Pd(1.2)/Co$_{90}$Fe$_{10}$(0.2)]$_3$/MgO(2)/[Co$_{90}$Fe$_{10}$(0.2)/Pd(1.2)]$_{10}$/cap-layer and buffer-layer/Pd(3.6)/Co$_{20}$Fe$_{60}$B$_{20}$(1.8)/MgO(2)/Co$_{20}$Fe$_{60}$B$_{20}$(1.8)/Pd(12)/cap-layer. All junctions were fabricated using a conventional photolithography technique with post-baking process of photoresist at 120°C. The MTJs were annealed at 200 ~ 300°C for 1 h under an out-of-plane magnetic field of 4 kOe. The TMR ratio was measured at RT using a dc four probe method with out-of-plane field of up to 8 kOe. The structures were investigated by high resolution transmission electron microscopy (HRTEM) and by fast Fourier transform (FFT) of the digitized HRTEM image. The compositions were analyzed by secondary ion mass spectrometer (SIMS) using Ar ion beam.

Figure 1 shows the annealing temperature ($T_a$) dependence of the TMR ratio for the MTJs with [Pd(1.2)/Co$_{90}$Fe$_{10}$(0.2)]$_3$/MgO(2)/[Co$_{90}$Fe$_{10}$(0.2)/Pd(1.2)]$_{10}$ and [Co$_{90}$Fe$_{10}$(0.2)/Pd(1.2)]$_3$/Co$_{20}$Fe$_{60}$B$_{20}$(1.8)/MgO(2)/Co$_{20}$Fe$_{60}$B$_{20}$(1.8)/[Pd(1.2)/Co$_{90}$Fe$_{10}$(0.2)]$_{10}$ multilayer stack structures. By inserting the CoFeB layers between CoFe/Pd multilayers and MgO barrier, the TMR ratio at RT increased from 1.5% to 43%. The TMR ratio of the MTJ with CoFeB insertion increased after annealing at 200°C and then rapidly decreased at $T_a$ over 250°C. In this MTJ system, no high TMR ratios of several hundred % shown in the previous reports of CoFeB/MgO/CoFeB MTJs with in-plane magnetic anisotropy[15,16] were obtained.

To understand the reasons for the low TMR ratio in the MTJs with CoFeB insertion layers, HRTEM was employed for structural characterization. Figure 2(a) shows cross-sectional HRTEM image for the MTJ with $Co_{20}Fe_{60}B_{20}$ insertion annealed at 300°C which showed low TMR ratio of less than 1%. The MgO barrier have (001) oriented texture, whereas the top and bottom CoFeB electrodes consist of fcc(111) or bcc(011) oriented texture with lattice fringe spacing of 0.210-0.218 nm according to the FFT images (not shown). In the partial area in which the lattice fringe in the HRTEM image was clearly observed, we could confirm the epitaxy of fcc-Pd(111)[110]//bcc-CoFe(B)(110)[111], as indicated by the lattice fringes with intersection angles of 71°+55° and 60°.[17,18] The fcc(111) or bcc(011) oriented crystallization of the inserted, initially amorphous, CoFeB layers is likely to be one of the reasons for the low TMR ratios, because high TMR ratios require bcc(001) oriented ferromagnetic electrodes and MgO (001) barrier. It is known that NiFe[19,20] or CoFe[21,22] adjacent to CoFeB acts as a template for crystallization of initially amorphous CoFeB into fcc(111) or bcc(011) oriented texture through annealing over crystallization temperatures which strongly depend on the thermal diffusion of B into the adjacent layers. We investigated the B diffusion by SIMS analysis. A simplified structure of $Pd(3.6)/Co_{20}Fe_{60}B_{20}(1.8)/MgO(2)/Co_{20}Fe_{60}B_{20}(1.8)/Pd(12)$ was used for this analysis. Figures 2(b) and 2(c) show the composition depth profiles of B, Pd and Co in the samples before and after annealing at $T_a$ = 300°C, respectively. In as-deposited state, B is located in CoFe, whereas after an-nealing B diffuses into Pd. These observations suggest a scenario that the diffusion of B into Pd layers reduces the crystallization temperature of CoFeB at the CoFeB/Pd interfaces, and crystallization of CoFe(B) into fcc(111) or bcc(011) oriented texture starts from the fcc(111) Pd seed layers. In the SIMS depth profiles before and after annealing, the high intensity of B is detected in MgO barrier. However, electron energy-loss spectroscopy analysis showed a large amount of B existed in the metal layers (buffer and cap layers) adjacent to the CoFe(B), and the concentration of B in the MgO barrier was low.[18,23] The difference in the B detection may be caused by the matrix effect in SIMS.

The above results strongly suggest that even higher TMR ratio may be realized once the Fe(Co) electrodes adjacent to the MgO barrier can be made bcc(001). We thus examined the annealing dependence of TMR ratio for the MTJs with Fe layers; note that Fe becomes bcc(001) when deposited on highly oriented (001) MgO. Figure 3 (a) shows the $T_a$ dependence of the TMR ratio for the MTJs with $[Co_{90}Fe_{10}(0.2)/Pd(1.2)]_3$/FM(1.8)/MgO(2)/FM(1.8)/$[Pd(1.2)/Co_{90}Fe_{10}(0.2)]_{10}$ (FM=$Co_{20}Fe_{60}B_{20}$ or Fe) stack structures along with the one shown in Fig. 1 for reference. The MTJs with Fe/MgO/Fe stack show lower TMR ratio (= 9.3 %) than those of the MTJs with $Co_{20}Fe_{60}B_{20}$/MgO/$Co_{20}Fe_{60}B_{20}$ stack. On the other hand, the TMR ratio of the $Co_{20}Fe_{60}B_{20}$/MgO/Fe stack increased and reached 67%. In the former, Fe is deposited on fcc(111) Pd and thus it is likely that Fe is in a crystal structure or orientation other than bcc(001), resulting in low TMR. On the other hand, in the case of $Co_{20}Fe_{60}B_{20}$/MgO/Fe stack, the top Fe layer deposited on MgO has a bcc(001) oriented texture as observed in the TEM images (not shown). Thus the enhancement of TMR ratio for the $Co_{20}Fe_{60}B_{20}$/MgO/Fe MTJs as annealing proceeds can be attributed to the (001) oriented top-Fe layer on MgO barrier; TMR increases as the bottom electrode/MgO interface becomes crystalline upon annealing.

In order to fully evaluate the TMR properties, it is necessary to stably realize parallel and antiparallel magnetization configurations. We tried to obtain a clear difference in coercivity between top and bottom ferromagnetic electrodes, i.e., the stable antiparallel state, by increasing the perpendicular magnetic anisotropy of the bottom CoFe/Pd multilayer electrode with *in-situ* annealing (see Fig. 4(a)). *In-situ* annealing at 350$^{o}$C for 1 h in vacuum chamber was applied right after depositing the bottom [CoFe/Pd]$_3$ multilayer, and then the CoFeB/MgO/Fe/[Pd/CoFe]$_{10}$ top stack structure was deposited without air exposure. Figure 4(b) shows the TMR ratio as a function of resistance-area product (*RA*) in MTJs with/without 350$^{o}$C *in-situ* annealing, which were annealed at $T_a$=200$^{o}$C. The MTJs without *in-situ* annealing have a wide distribution of TMR ratio. In contrast, the TMR ratio for the MTJs with *in-situ* annealing was enhanced to 78% together with reduction of its distribution. A typical TMR loop is shown in Fig. 4(c). The improved distribution of TMR ratio is consistent with the above-mentioned scenario regarding stabilization of the antiparallel magnetization configuration by *in-situ* annealing.

In summary, we investigated the TMR properties and film structures of perpendicular MTJs with CoFe/Pd multilayer and different insertion layers such as CoFeB and Fe. The insertion of CoFeB layers between CoFe/Pd multilayers and MgO barrier resulted in an increase of TMR ratio from a few % to up to 43%. By applying combination of bottom CoFeB and top Fe insertion layers, the TMR ratio reached 67%. For the MTJs with *in-situ* annealing The TMR ratio of 78% was observed. However, the TMR ratio rapidly decreased at $T_a$ over 250°C. The degradation of the TMR ratio for the MTJs annealed at high $T_a$ may be related to the crystallization of CoFe(B) into fcc(111) or bcc(011) texture resulting from the diffusion of B into Pd layers. Even higher TMR ratio may be expected in MTJs with perpendicular magnetic CoFe/Pd multilayer electrodes, if one can realize bcc(001) orientation of CoFeB, for example, by suppression of B diffusion between the CoFe/Pd multilayer and the CoFeB layer.

This work was supported in part by the "High-Performance Low-Power Consumption Spin Devices and Storage Systems" program under Research and Development for Next-Generation Information Technology of MEXT. The authors wish to thank Y. Ohno for discussion and I. Morita and T. Hirata for their technical support in MTJ fabrication and discussion.

Figure captions

FIG. 1. TMR ratio as a function of annealing temperature ($T_a$) for the MTJs with [Pd(1.2)/Co$_{90}$Fe$_{10}$(0.2)]$_3$/MgO(2)/[Co$_{90}$Fe$_{10}$(0.2)/Pd(1.2)]$_{10}$ and [Co$_{90}$Fe$_{10}$(0.2)/Pd(1.2)]$_3$/Co$_{20}$Fe$_{60}$B$_{20}$(1.8)/MgO(2)/Co$_{20}$Fe$_{60}$B$_{20}$(1.8)/[Pd(1.2)/Co$_{90}$Fe$_{10}$(0.2)]$_{10}$ multilayer stacks. TMR ratio at $T_a$ = 120$^\circ$C was obtained in as-made MTJ with post-baking process of photoresist at 120$^\circ$C.

FIG. 2. (a) Cross-sectional HRTEM image for a buffer-layer/Pd(1.2)/[Co$_{90}$Fe$_{10}$(0.2)/Pd(1.2)]$_3$/Co$_{20}$Fe$_{60}$B$_{20}$(1.8)/MgO(2)/Co$_{20}$Fe$_{60}$B$_{20}$(1.8)/[Pd(1.2)/Co$_{90}$Fe$_{10}$(0.2)]$_{10}$/Pd(1.2)/cap-layer stack after annealing at 300$^\circ$C. Composition depth profiles of B, Pd, and Co in the samples with buffer-layer/Pd(3.6)/Co$_{20}$Fe$_{60}$B$_{20}$(1.8)/MgO(2)/Co$_{20}$Fe$_{60}$B$_{20}$(1.8)/Pd(12)/cap-layer stack (b) before and (c) after annealing at 300$^\circ$C analyzed by SIMS.

FIG. 3. TMR ratio as a function of annealing temperature ($T_a$) for the MTJs with buffer-layer/Pd(1.2)/[Co$_{90}$Fe$_{10}$(0.2)/Pd(1.2)]$_3$/FM(1.8)/MgO(2)/FM(1.8)/[Pd(1.2)/Co$_{90}$Fe$_{10}$(0.2)]$_{10}$/Pd(1.2)/cap-layer stack where FM is Co$_{20}$Fe$_{60}$B$_{20}$ or Fe insertion.

FIG. 4. (a) $M$-$H$ loops for the samples consisting of buffer-layer/Pd(1.2)/ [Co$_{90}$Fe$_{10}$(0.2)/Pd(1.2)]$_3$

(with/without in-situ annealing at 350$^o$C)/MgO(2)/Ta(5)/Ru(5). (b) TMR ratios as functions of resistance-area product (*RA*) for the MTJs with (filled circles) and without (open circles) *in-situ* anneal at 350$^o$C just after depositing bottom CoFe/Pd multilayer. (c) *R-H* curve at RT of the MTJ with *in-situ* anneal.

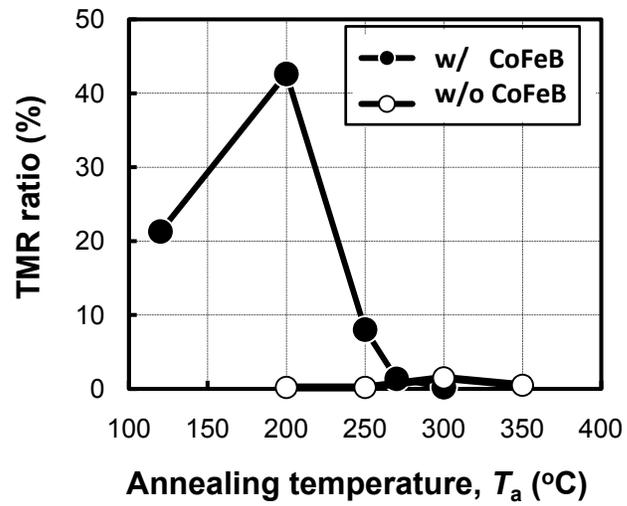

FIG.1

K. Mizunuma et al.

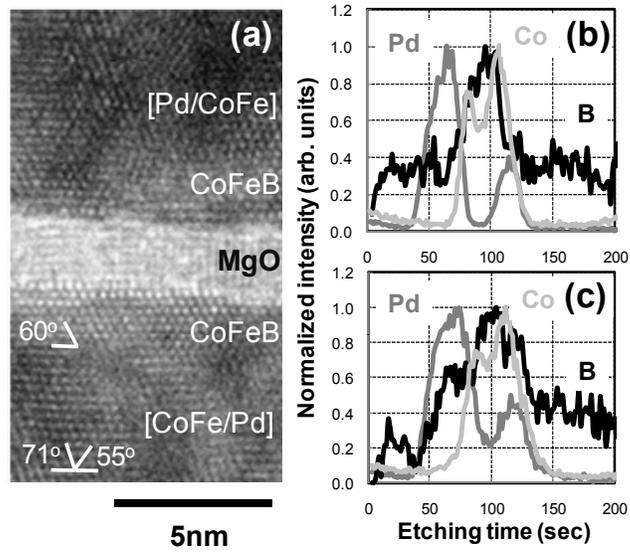

FIG.2

K. Mizunuma et al.

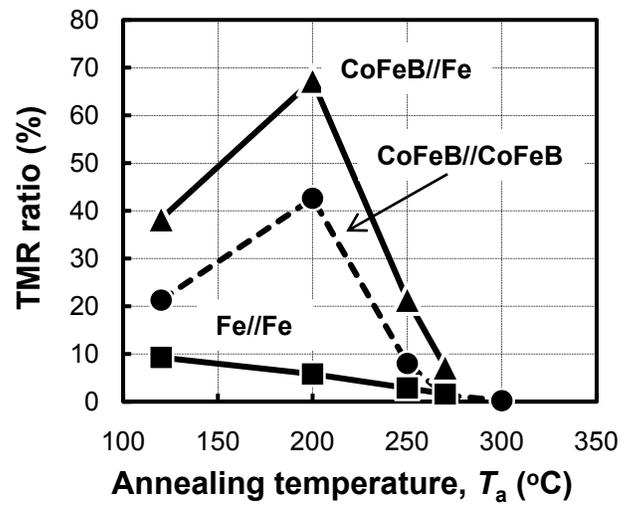

FIG.3

K. Mizunuma et al.

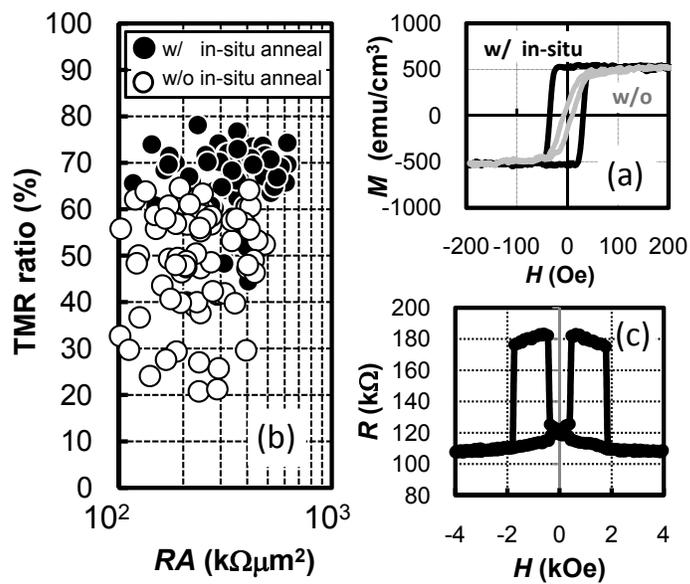

FIG.4

K. Mizunuma et al.